\documentclass[runningheads]{llncs}
\usepackage[title]{appendix}
\usepackage{graphicx}
\usepackage{caption}
\usepackage{subcaption}

\usepackage[dvipsnames]{xcolor}

\usepackage{algorithm}
\usepackage{algpseudocode}
\usepackage{breqn}
\usepackage{amssymb}

\newcommand{\pfirst}{GSD}

\begin{document}
\title{Drift Detection: Introducing Gaussian Split Detector}
\titlerunning{Gaussian Split Detectors}
%

\author{Maxime Fuccellaro \and
Laurent Simon \and
Akka Zemmari}

%
\institute{University of Bordeaux, CNRS, Bordeaux INP, LaBRI, UMR 5800,
351, cours de la Libération F-33405 Talence}
%
\maketitle              
\begin{abstract}
Recent research yielded a wide array of drift detectors. However, in order to achieve remarkable performance, the true class labels must be available during the drift detection phase. This paper targets at detecting drift when the ground truth is unknown during the detection phase. To that end, we introduce Gaussian Split Detector (\pfirst) a novel drift detector that works in batch mode. \pfirst~is designed to work when the data follow a normal distribution and makes use of Gaussian mixture models to monitor changes in the decision boundary. The algorithm is designed to handle multi-dimension data streams and to work without the ground truth labels during the inference phase making it pertinent for real world use. In an extensive experimental study on real and synthetic datasets, we evaluate our detector against the state of the art. We show that our detector outperforms the state of the art in detecting real drift and in ignoring virtual drift which is key to avoid false alarms.

\end{abstract}

\section{Introduction}


Few data distributions remain stationary over long periods of time. Machine learning models are constructed based on the assumption that the data encountered during the training stage and the future inference data share the same distribution. When models are trained and then deployed for inference, a change in the underlying data distribution can cause a drop in performance \cite{kelly1999impact}. This shift in distribution is known as Concept Drift (CD) and detecting it will be the focus of this paper. The democratization of predictive modeling has made concept drift an active research topic as it cripples in production systems. Concept drift must not be confused with anomaly detection where few samples are out of distribution. Drift is not domain specific and impacts every field including text and video \cite{rabanser2019failing}.

\textit{Real drift} is a change in distribution that modifies the decision boundary. Real drift impacts a model's ability to predict instances classes. When real drift occurs, a drop of a model's predictive performances is systematically observed. \textit{Virtual drift} is a change of distribution that does not affect the decision boundary. Since virtual drift does not impact the model's performance, we don't believe its occurrence warrants the re-training of the model. We believe a good drift detector should detect real drifts while ignoring virtual drifts.

Assume $X$ is a set of variables used to predict the target class vector $y$. Three primary causes of concept drift are identified \cite{webb2016characterizing}: changes in the class distribution $\mathbb{P}(y \mid X)$, the feature space $\mathbb{P}(X \mid y)$ or class priors $\mathbb{P}(y)$. Drift of type $\mathbb{P}(y \mid X)$ happens when the decision boundaries change. Drift type $\mathbb{P}(X \mid y)$ happens when values reach unseen domains. Different drift speeds have been identified \cite{lu2018learning}. If the distribution change is sudden, it is referred to as \textit{abrupt drift}, whereas if the distribution shift is slow over time, it is called \textit{gradual drift}. When a distribution oscillates between two or more concepts, it is called \textit{recurrent drift}.



In this paper, we introduce a novel drift detection method, Gaussian Split Detector (\pfirst).
\pfirst~computes decision boundaries of the most informative features on the training set. During the inference phase, the Expectation Maximization \cite{dempster1977maximum} (EM) algorithm estimates the parameters of the Gaussian mixture distribution on the selected features. The parameters are used to compute the new decision boundaries. Drift detection is based on the difference of the training boundaries and the inference boundaries. As \pfirst~works in any number of dimensions, it does not need labels during inference and outperforms the state of the art in thorough experiments.

In Section 2, we present related work and position our paper. The \pfirst~algorithm is described and evaluated in Section 3. Section 4 concludes this paper.

\section{Related work}

Concept drift has become the focus of recent work in the last few years. Recent research yielded a wide array of models handling drift intrinsically. As well as drift detectors that work jointly with a predictive model. If the ground truth is available shortly after inference, some detectors manage to achieve almost perfect drift detection. Two hypothesis are usually made in supervised drift detection. The first is that recent data shares the same distribution as upcoming data. The second is that a change in a model's error rate is a strong indicator of the presence of real drift.


With the assumption that recent and upcoming data share the same distribution, one approach is to continuously update a pool of models. In \cite{elwell2011incremental} a batch of new data is evaluated by a pool of models, and the contributions of each individual model are weighted based on its recent performance. With each new batch of data, a model is trained on it and added to the pool. Consistently poorly performing models are removed. This ensures quick adaptation to recurring drifts while keeping a strong level of performance. Techniques that operate on a dynamic pool of models have been extensively studied in \cite{brzezinski2013reacting,kolter2007dynamic}.

Detection methods based on the second hypothesis have received significant attention since they can reliably identify genuine drift while ignoring virtual drift. In \cite{baena2006early} and \cite{de2018conceptss} a model's error rate is monitored and a decline in performance is interpreted as drift. Various statistical tests \cite{nishida2007detecting} are employed to monitor the error rate and signal the occurrence of drift. These methods can reliably identify real drift while disregarding virtual drift as it does not impact a model's performances.

Updating a pool of models and monitoring the error rate are both effective techniques for handling drift. Both approaches require immediate access to the true labels after the inference stage. This is a very strong hypothesis for real-world situations where class labels may be never known \cite{hinder2023change}. Unsupervised methods have been the focus of research to deal with ground truth unavailability.

To detect drift based on the feature space instead of the true labels, window-based techniques have been studied. In \cite{bifet2007learning} the authors introduce ADWIN that maintains a reference window consisting of past instances. The window size dynamically adjusts based on whether drift is detected. Specifically, when no change is detected, the window expands, while in the presence of drift, the window rapidly contracts. The detection mechanism of ADWIN involves repeatedly separating the window into two smaller windows based on the observation's age. Drift is detected when the averages of the values of these sub-windows exceed a threshold. Other window-based detectors have been proposed such as in \cite{raab2020reactive}.

Rather than relying on one-dimensional sliding windows, some methods detect drift on all the feature space. In \cite{frittoli2021change}, the authors introduce QT-EWMA where concept drift is detected using an exponential moving average to monitor the distribution of a QuantTree histogram built on the training data. The authors in \cite{heusinger2020reactive} use minimum enclosing balls to identify abrupt or gradual changes. A ball is defined as a centroid and a minimum radius that encompasses all the samples in the ball. If a significant number of values are identified as outliers, a drift is detected and the centroid is updated to fit the latest concept.

Some authors use the hypothesis that if past and present data can be sorted, there is a drift.
In \cite{gozuaccik2019unsupervised}, the authors train a model to distinguish between past and recent data. To prevent the model from identifying this distinction based on obvious timestamp-like aggregates, these were removed before training the model. The researchers evaluated the model's performance using the AUC metric. In \cite{black2003learning}, the authors use time as a means to detect drift by incorporating the timestamp attribute in the observations and training a model to predict the target variable for both past and recent data. If the timestamp attribute is considered as an informative feature, then the target variable is dependent on time, which indicates presence of a drift.

In \cite{cerqueira2020unsupervised}, the authors propose a different unsupervised approach to detect drift. They train a teacher model on past labeled data, and then train a second model, known as the student model, to mimic the behavior of the teacher model. During the inference phase, the authors monitor the error rate of the student model and use the method presented in \cite{bifet2007learning} to detect a drift if the error rate exceeds a certain threshold.

The authors in \cite{rabanser2019failing} investigate drift detection in the context of unsupervised image classification. The dimension is first reduced, followed by a two-sample test to identify drift. They evaluate a variety of dimension reduction and two-sample tests, including MMD \cite{gretton2012kernel} and KS, by applying various types of shifts to images. In \cite{castellani2021task}, the authors propose an approach where the target class is incorporated in the dimension reduction mechanism, allowing the detector to ignore virtual drift.



The idea behind concept drift detection by statistical tests is that a distribution change will be a strong indicator of drift \cite{heusinger2020reactive,rabanser2019failing}. A distribution change will enable the detection of drift, but will not discriminate between real drift and virtual drift. To the best of our knowledge, few algorithms are able to discriminate between real drift and virtual drift in a multivariate setting without access to true labels after detection such as Discriminative Drift Detector \cite{gozuaccik2019unsupervised}, QT-EWMA \cite{frittoli2021change}, Student-Teacher (ST) \cite{cerqueira2020unsupervised}, Task Sensitive Drift Detector (TSDD) \cite{castellani2021task}.

In this paper we introduce a novel algorithm to detect drift: Gaussian Split Detector (\pfirst). \pfirst~is designed to handle binary classification problems. GSD successfully detects real drifts while triggering very few false positive by ignoring virtual drifts. 
The detector does not need true class labels to operate during the inference phase.

\section{Gaussian Split Detector}

In this Section we introduce \pfirst, an algorithm that monitors the decision boundaries shifts using a Gaussian mixture model in order to detect drift. We simplify the assumption made by the authors of \cite{brzezinski2021impact} by making the hypothesis that the data distributions of any feature used for detection is a sum of two Gaussian distributions each corresponding to a class.

\begin{figure}
\centering
\includegraphics[width=0.99\textwidth, trim={4cm 1cm 3cm 3cm},clip]{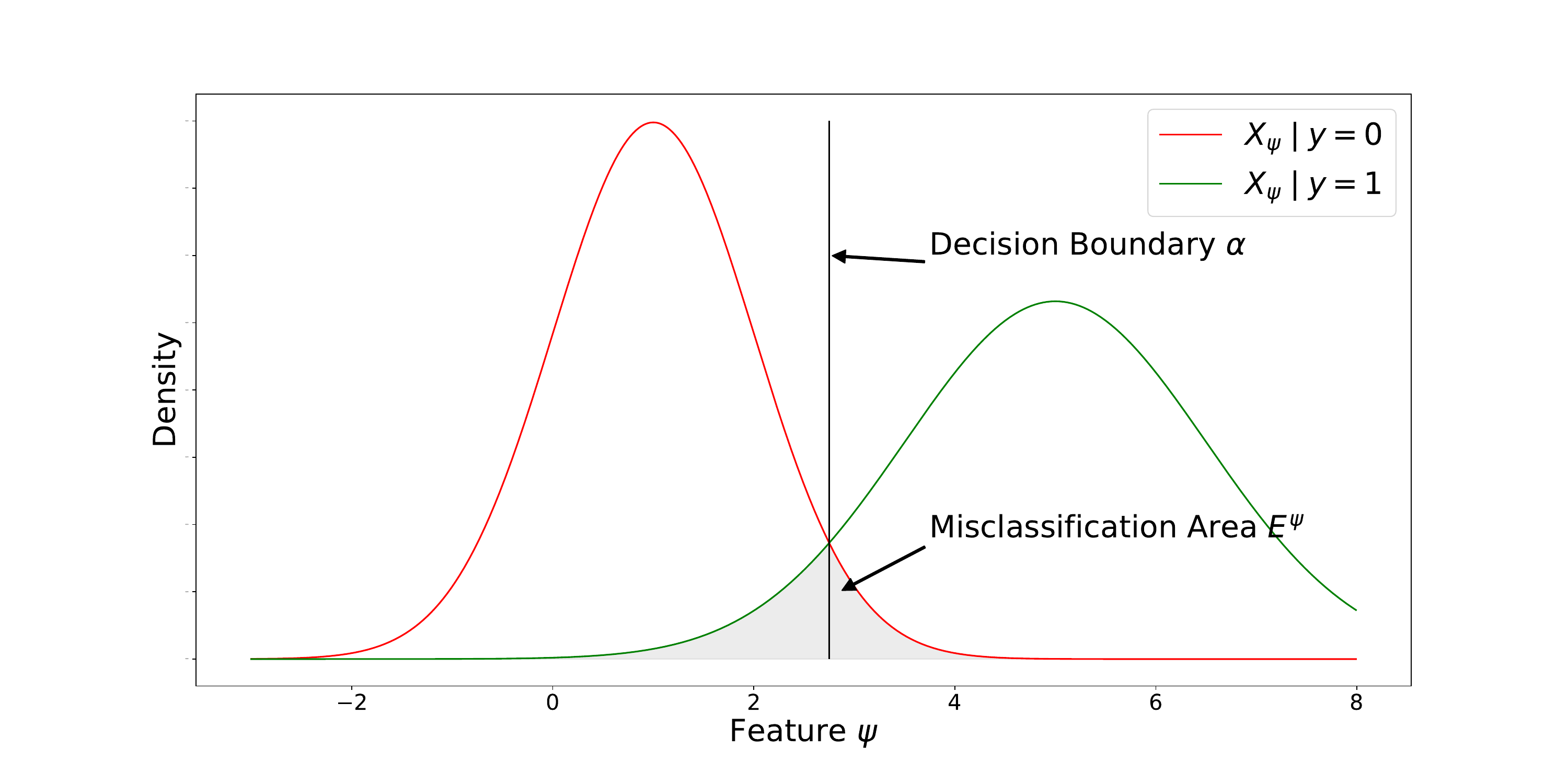}
\caption{In this figure, we plot the weighted density functions for the positive (in green) and negative (in red) samples. The decision boundary $\alpha$ is marked by the black line.
} \label{fig1}
\end{figure}

We start by building an ensemble of $n$ single \textit{Bayesian splits}. In a similar fashion to the random forest algorithm \cite{breiman2001random}, each split is randomly given a subset of samples and features.
During the training phase, for each feature $\psi$ available to the split, the mean $\mu^{\psi}_{c}$, variance $\sigma^{\psi}_{c}$ and proportion $p^{\psi}_{c}$ are calculated for each class $c$. The proportion is used to find the optimal decision boundary with regards to a possible class imbalance scenario, we have $p^{\psi}_{0} + p^{\psi}_{1} = 1$.
Since we assume the variables are drawn from a Gaussian distribution we use these parameters to find the decision boundary. It is the intersection that minimizes the misclassification error rate of the two Gaussian curves. The feature selected for the split is the one that has the smallest misclassification area $E^{\psi}$. Figure \ref{fig1} illustrates the process of computing the decision boundary.

During inference, the EM algorithm allows us to estimate the parameters of a Gaussian mixture distribution. The parameters are then used to compute a new decision boundary $\hat{\alpha}$. The decision boundary difference is used to detect drift.

\subsection{Mathematical formulation}

\subsubsection{Hypothesis}

We assume that the predictive task is binary classification.

Furthermore, for each variable $\psi$, we make the hypothesis that:
\begin{equation}\label{eq:hyp1}
\forall c \in \{0, 1\} : X_{\psi} \mid y=c \sim \mathcal{N}(\mu^{\psi}_{c}, \sigma^{\psi}_{c})
\end{equation}
where $X_{\psi} \mid y=c$ denote the samples for feature $\psi$ labelled as belonging to class $c$. \pfirst~does not require the features \textit{to be independent}.

\subsubsection{Algorithm}

Let $F_{c}^{\psi}(x)$ be the cumulative distribution function of the estimated Gaussian distribution for the variable $\psi$ of the samples of class $c$. Let $p_i^{\psi}$ be the proportion of samples of class $i$. Let $\alpha$ denotes the decision boundary that minimizes the misclassification area between the two classes. The misclassification area is calculated with equation \ref{eq:error}.

\begin{equation}\label{eq:error}
\begin{gathered}
    E^{\psi} = \min \biggl( \min \left[ p^{\psi}_{0} \times F_{0}^{\psi}(\alpha), p^{\psi}_{1} \times F_{1}^{\psi}(\alpha) \right] +\\
    \min \left[ p^{\psi}_{0} \times(1-F_{0}^{\psi}(\alpha)), p^{\psi}_{1} \times (1-F_{1}^{\psi}(\alpha)) \right] \biggl)
\end{gathered}
\end{equation}

As in classic decision trees, the feature used in the split minimizes the misclassification probability across all features. For each feature $\psi$, the optimal boundary, $\alpha$, is one of two intersections of the weighted Gaussian distributions. $\alpha$ can be computed analytically by solving a quadratic equation where the coefficients are:

$$a =  \frac{1}{2(\overline{\sigma_0^{\psi}})^2} - \frac{1}{2(\overline{\sigma_1^{\psi}})^2}, b = \frac{\overline{\mu^{\psi}_1}}{(\overline{\sigma_1^{\psi}})^2} - \frac{\overline{\mu^{\psi}_0}}{(\overline{\sigma_0^{\psi}})^2}$$

$$c = \frac{(\overline{\mu_0^{\psi}})^2}{2(\overline{\sigma_0^{\psi}})^2} - \frac{(\overline{\mu_1^{\psi}})^2}{2(\overline{\sigma_1^{\psi}})^2} - \ln \biggl(\frac{\overline{p^{\psi}_0}\,\overline{\sigma^{\psi}_1}}{\overline{p^{\psi}_1}\,\overline{\sigma^{\psi}_0}}\biggr)$$


The $\alpha$ chosen minimizes the misclassification area. In some rare cases, the distribution functions do not intersect. This happens when the means are close and there is a steep class imbalance which causes a weighted distribution to encompass another. When encountered, the feature is dropped. We use here an ensemble of single splits for two reasons:
\begin{itemize}
    \item In order to properly estimate the parameters of the Gaussian distribution we need a large enough sample size. This is not achievable when the data is partitioned by a large number of splits.
    \item It allows the model to work when the variables are not independent. In that case, a shift in the first split may drastically change the distribution of the final partitioning when we update its decision boundary.
\end{itemize}

\subsection{Overview of the detection phase}

\begin{algorithm}
\caption{\pfirst~- Inference}\label{alg:cap}
\begin{algorithmic}
\State \textbf{Parameters:}
\State - $\tau$ : Minimum ratio of drifting tree to flag drift
\State - $\beta$ : Vector of thresholds to flag a drift
\State \textbf{Inputs:}
\State - $\mathcal{M} = \{\mathcal{M}_0, ..., \mathcal{M}_k\} $ : Trained ensemble of $k+1$ single Gaussian splits
\State - $\alpha = \{\alpha_0, ..., \alpha_k\}$ : Decision boundaries of the Gaussian splits
\State - $I \in \mathbf{R^{m \times d}}$ : Test set with $\mathbf{d}$ features and $\mathbf{m}$ samples

\State \textbf{Variables:}
\State - $\gamma = 0$ : Number of drifting splits
\end{algorithmic}

\begin{algorithmic}[1]
\For{all $\psi$ used in $\mathcal{M}$}
    \State Run EM on $I[\psi]$ and get $\hat{\mu^{\psi}_i}, \hat{\sigma^{\psi}_i}, \hat{p^{\psi}_i}, i \in \{0, 1\}$
    \State Get the Gaussian decision boundary $\hat{\alpha}_{\psi}$
    \If{$|\alpha_{\psi}-\hat{\alpha}_{\psi}| \geq \beta_{\psi}$}
    \State $\gamma = \gamma + 1$
    \EndIf
\EndFor
\If{$\frac{\gamma}{k+1} \geq \tau $}
\State A drift is triggered
\EndIf

\end{algorithmic}
\end{algorithm}

The drift detection phase is formally described in Algorithm \ref{alg:cap}. Let $\mathcal{M}$ be the ensemble of Gaussian splits computed during the training phase. In lines 1-2, for each feature in $\mathcal{M}$ we run the EM algorithm on the inference data. This gives the estimated parameters ($\hat{\mu_c^{i}}, \hat{\sigma_c^{i}}, \hat{p_c^{i}}$) for each subsequent feature $i$ and class $c$ of the two Gaussian distributions that constitute the overall distribution.
In line 3, we find the new decision boundary $\hat{\alpha}_{\psi}$ by solving (\ref{eq:error}). It is the intersection of the two weighted Gaussian probability density functions. When the EM algorithm does not converge, the tree's output is not taken into account.

The $\beta$ vector is used to determine if a change in the decision boundary in between the training data and inference data is high enough to signal a drift. To compute it, we split the labelled data into a train set containing 75\% of the instances and a validation set. We then proceed to build a Gaussian split on all features of the training set. We then isolate the two components of the Gaussian mixture with the EM algorithm on the validation set as it is done in the detection phase. With $\hat{\alpha}^i$ the estimated decision boundary on the validation data for feature $i$, the $\beta$ vector is defined as $[|\alpha^0-\hat{\alpha}^0|, ..., |\alpha^d-\hat{\alpha}^d|]$.  

In lines 4 through 6, when the difference in the decision boundary exceeds the feature dependant $\beta_\psi$ parameter, the $\gamma$ count is incremented. In line 8, if the ratio of drifting split exceeds a user defined threshold $\tau$, a drift is triggered. The $\tau$ value controls the sensibility of the detector. A low value will likely trigger false alarms while a high value might lead the detector to miss some real drifts.

\subsection{Experimental results}

\begin{figure}
     \centering
     \begin{subfigure}[b]{0.49\textwidth}
         \centering
         \includegraphics[width=\textwidth]{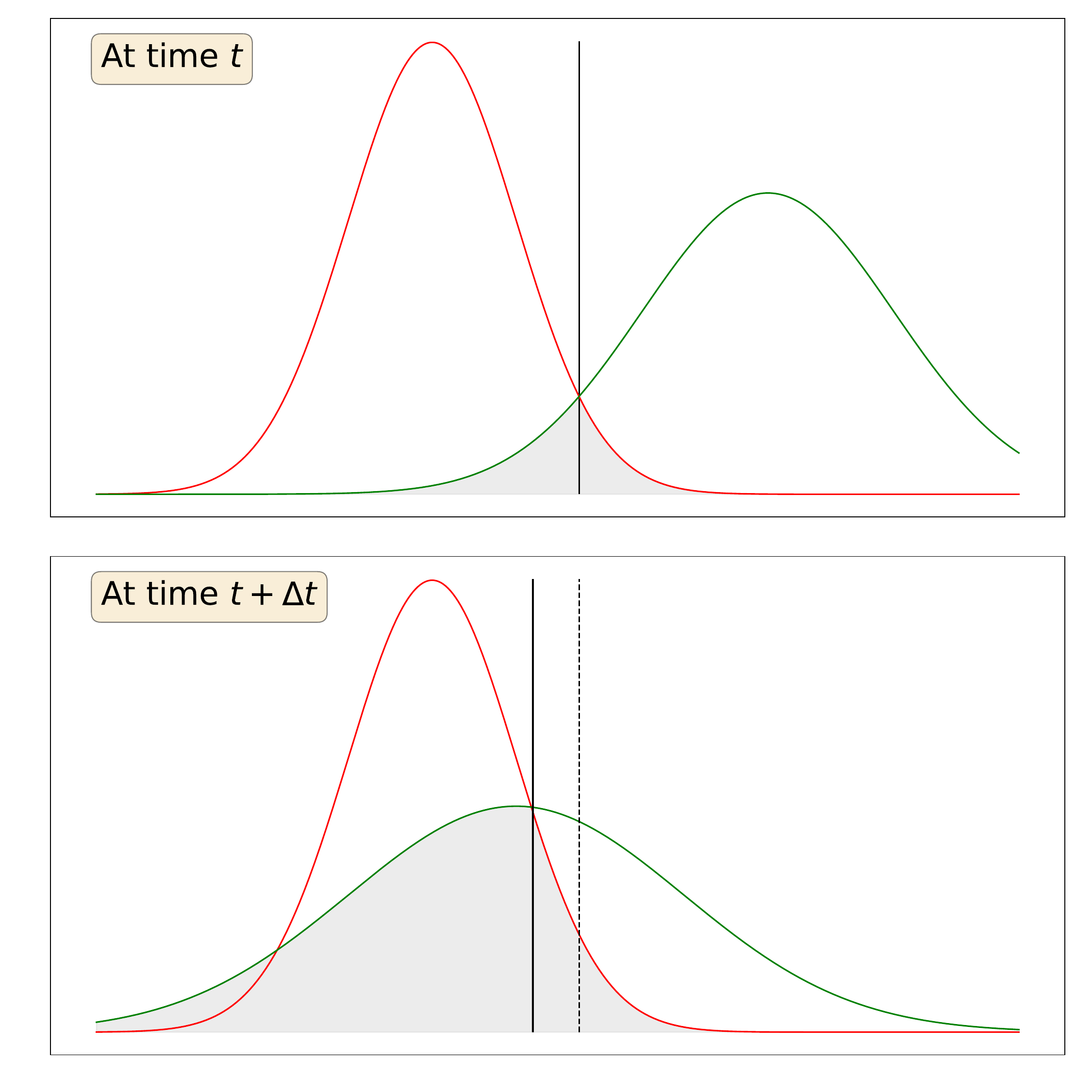}
         \caption{Real drift}
         \label{fig:threegraphsreal}
     \end{subfigure}
     \hfill
     \begin{subfigure}[b]{0.49\textwidth}
         \centering
         \includegraphics[width=\textwidth]{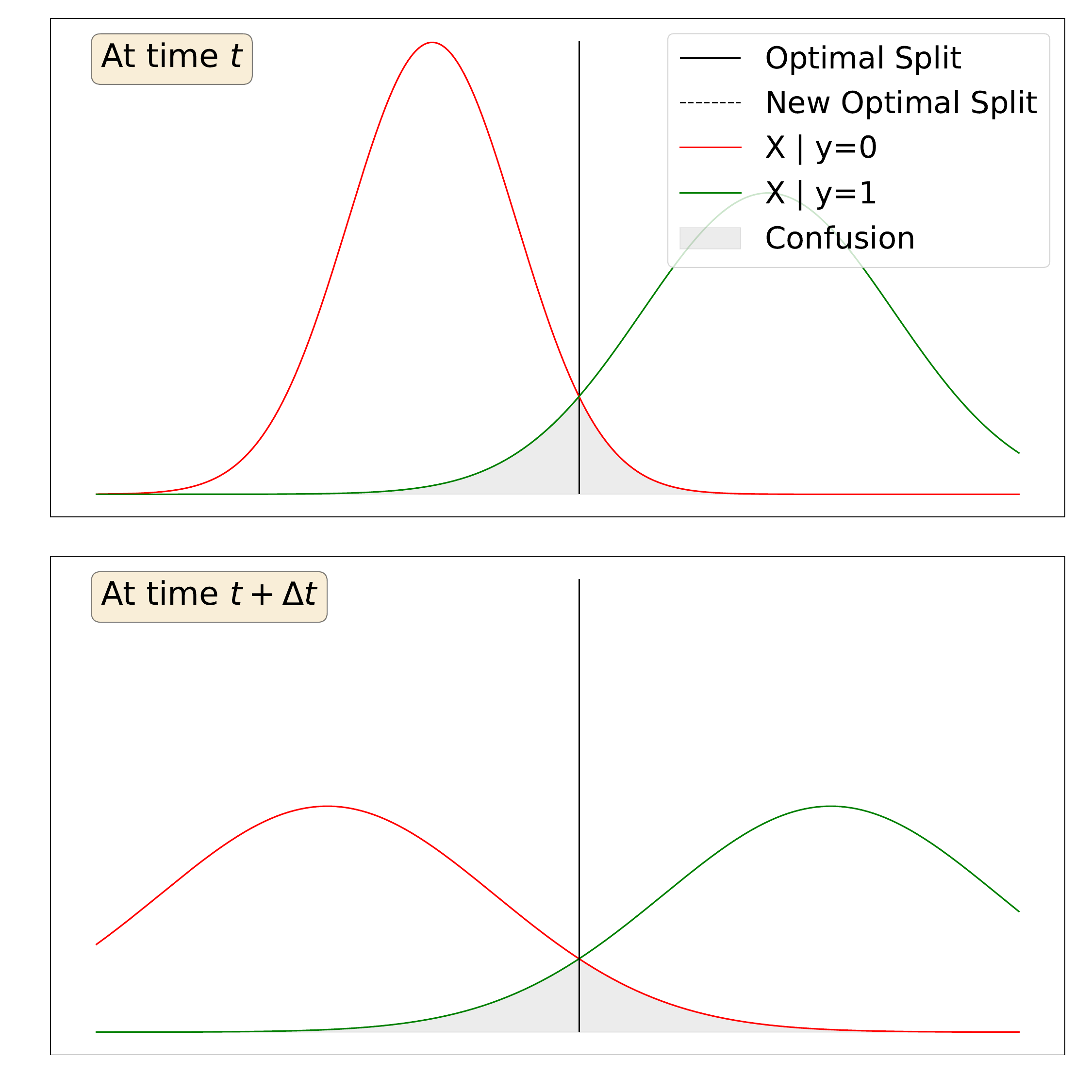}
         \caption{Virtual drift}
         \label{fig:threegraphsvirtual}
     \end{subfigure}
        \caption{Illustration of real and virtual drift}
        \label{fig:threegraphs}
\end{figure}

In Figure \ref{fig:threegraphs} we show how real drift affects the decision boundary and how the decision boundary can remain unchanged with virtual drift. In Figure \ref{fig:threegraphsreal}, real drift is illustrated as the optimal split shifts with the distribution change. In Figure \ref{fig:threegraphsvirtual}, virtual drift is shown as the decision boundary remains unchanged despite of the distribution change.
We evaluate \pfirst~against a panel of five state of the art detectors on both real and synthetic datasets. In order to test our method's ability to detect real drift and ignore virtual drift, perturbations are made on the datasets. The drift induction experimental procedure used in this section is standard when testing drift detectors \cite{castellani2021task,sethi2017reliable,sethi2016monitoring}.

\subsubsection{Experimental protocol}
The usual procedure to test algorithms suited to handle drift when true class labels are available after inference, is the test-then-train approach. A model predicts the class on a batch of samples, then, the true class is revealed and the model updates itself. The global prediction accuracy is then used to rank models.

This setup is not suited for models that do not rely on true labels availability. In most datasets used to benchmark drift handling methods, the presence of drift is only assumed or artificially introduced by sorting the observations on an attribute. To know the exact drift occurrence and its impact, two distinct perturbations are introduced. The \textit{Step Drift}, where a subset of the features are shuffled and the \textit{Noise Drift}, where Gaussian $\mathcal{N}(1, 1)$ noise is added to a subset of features (Gaussian noise with mean equal to 1 is used to change the mean of the distribution, not to obfuscate the signal).
The idea is now to be able to artificially generate real drift and virtual drift. To create virtual drift, we add one of the two perturbations on the 25\% \textit{least} important features as to their predictive power of the class labels. In doing so, we hope that changing the distribution of several features will not affect a predictive model's performances. To create real drift, we modify the 25\% \textit{most} informative features by adding one of the two perturbations. The intuition is that a change of distribution on the most important features is likely to cause a drop of performance in a predictive model. To find 25\% most and least important features, we train a Random Forest Classifier over the training data. We choose this model as it is a robust, widely used model that achieves good level of performance on the datasets. For each dataset, we introduce the noise and step perturbations on the most and least sets of features thus creating four distinct sets : Least Noise, Least Step, Most Noise and Most Step.

In order to have stationary non-drifting data before adding our generated drifts, we first randomly shuffle the observations. Each dataset is then randomly partitioned into three subsets: a train set with 60\% of the data, a validation set and a drift set each containing 20\% of the data. Four distinct copies of the drift set are independently modified with the four different perturbations described above. In order to assess if a drift is virtual or real, we fit a Random Forest Classifier on the train set before reporting its accuracy on the train set, validation set and the four different drift sets. The drop of the model's accuracy between the different sets is used to classify drift as virtual or real. If the difference in accuracies between the validation set and the training set is lower than that of the validation set and the drift set, we consider the drift induced to be real, otherwise, it is considered as a virtual drift. We apply this protocol on all datasets except for the Insect dataset \cite{souza2020challenges}, where the exact abrupt drift points are known. As \pfirst~only works on continuous features, categorical features were removed from the datasets. 

 We evaluate the performance of \pfirst~against other state of the art unsupervised drift detectors :
\begin{itemize}
    \item ADWIN \cite{bifet2007learning} with $\delta=0.7$
    \item Discriminative Drift Detector (D3) \cite{gozuaccik2019unsupervised}
    \item QT-EWMA \cite{frittoli2021change}
    \item Student-Teacher (ST) \cite{cerqueira2020unsupervised}
    \item Task Sensitive Drift Detector (TSDD) \cite{castellani2021task}
\end{itemize}

All detectors were used with their default parameter values except for ADWIN, where the $\delta$ value was optimized with a grid search as the default value of $0.002$ did not showcase good results.


We evaluate our detector on the real datasets: Adult \cite{hettich1999uci}, Bank, Chess \cite{vzliobaite2011combining}, CoverTypes \cite{bifet2010moa}, Elec \cite{gama2014survey}, Digits 08, Musk used in \cite{sethi2017reliable}, Insect, Luxembourg \cite{jowell2003central,vzliobaite2011combining}, Wine \cite{hettich1999uci} and on the synthetic datasets: Hyperplane \cite{domingos2001mining} and Waveform \cite{bifet2010moa}. 

\begin{table}[h!]
\caption{Overview of the dataset used in our experiment. All but one RW dataset are binary classification problems. 2 RW datasets contain more than 100 features. For the synthetic datasets, we limit the number of generated observations at 10 000.}
\centering
\label{tab:datasets}
\begin{tabular}{c|c|c}

Dataset & Dimensions & Classification \\
\hline
Adult &   (48842, 66) &   Binary \\
Bank &   (45211, 49) &   Binary \\
Chess & (534, 7) & Binary \\
Cov &  (110393, 51) &   Multi-class (7) \\
D08 &    (1499, 17) &   Binary \\
Elec &   (45312, 15) &   Binary \\
Lux. & (1901, 20) &   Binary  \\
Musk &   (6598, 167) &   Binary \\
Wine &    (6497, 13) &   Binary \\
\hline
Hyp. &   (10000, 11) &   Binary \\
Wav. &   (10000, 41) &   Multi-class (3) \\
\hline
\end{tabular}
\end{table}


\begin{table}
\caption{Real and virtual drift classification}\label{tab:virtualone}
\centering
\label{tab:driftType}
\begin{tabular}{c|cc|cccc}
{} &  Train &  Validation &  Least Noise &  Least Step &  Most Noise &  Most Step \\
\hline
Adult &   1. &        .85 &         .85 &        .85 &        \textbf{.28} &       \textbf{.59} \\
Bank  &   1. &        .94 &         .92 &        .94 &        \textbf{.52} &       \textbf{.52} \\
Chess &   1. &        .78 &         .77 &        .76 &               .77   &       \textbf{.37} \\
Cov   &   .99 &        .85 &         .84 &        .84 &        \textbf{.49} &       \textbf{.46} \\
D08   &   1. &        1. &         .99 &        .99 &        \textbf{.77} &       \textbf{.69} \\
Elec  &   1. &        .89 &         .87 &        .87 &        \textbf{.57} &       \textbf{.62} \\
Lux. & 1. & 1.         &        1.0  &        1. &        \textbf{.55} &       \textbf{.51} \\
Musk  &   1. &        .99 &         \textbf{.94} &        \textbf{.97} &        \textbf{.51} &       \textbf{.56} \\
Wine  &   1. &        1. &         1. &        1. &        \textbf{.63} &       \textbf{.44} \\
\hline
Hyp.  &   1. &        .87 &         .87 &        .85 &        \textbf{.71} &       .87 \\
Wav.  &   1. &        .85 &         .85 &        .85 &        .72 &       \textbf{.41} \\
\hline
\end{tabular}
\end{table}

In Table \ref{tab:virtualone}, we show the effects that the four different types of perturbations on the test set have on the accuracy. In bold are the settings that produce real drift. We note that the Musk dataset exhibits real drift for all types of perturbations. In order to correctly conduct our experiment, each experiment is performed over 10 runs.

We group the experimental results in two tables (Table \ref{tabvirtualone} and Table \ref{tabrealone}). Table \ref{tabvirtualone} reporting the performance of detectors when shown virtual drift, Table \ref{tabrealone} reporting the results on real drift.


\subsubsection{Results}

Table \ref{tabvirtualone} evaluates the detectors on their power to ignore virtual drift by reporting the false positive rate. A low false positive rate signifies that the detector accurately ignored virtual drift. \pfirst, TSDD and ST consistently ignore the virtual drifts. \pfirst~makes no false positive detection in seven datasets, while keeping the false positive rate under 50\% in four others. TSDD accurately ignores the virtual on nine datasets. The student-teacher model successfully ignores virtual drift six times but classify virtual drift as real drift in four datasets. We see that our model ignores best virtual drift when the dimension is lower. ADWIN, QT-EWMA and D3 almost systematically flag virtual drift as real drift. D3 outperforms ADWIN and QT-EWMA on the Waveform dataset and on the Insect dataset, both with the highest number of features.

\begin{table}
\caption{Virtual Drift Detection}\label{tabvirtualone}
\centering
\begin{tabular}{lrrrrrrrrrrr}
\hline

 & Hyper. & Wave.  & Cov & Elec & Adult & Bank & D08 & Wine &    Ins. & Chess & Lux. \\
\hline
\textbf{\pfirst}  & \textbf{0.0} & 0.40 &  \textbf{0.0} &  \textbf{0.0} & 0.30 & \textbf{0.0} & \textbf{0.0} &  \textbf{0.0} &   0.32 & \textbf{0.0} & 0.45\\
\hline
ADWIN           &        0.33 &              0.73 &  0.5 &  0.5 &      1.0 &      1.0 &          1.0 &      1.0 &   1.0 &  0.76  &   1.0 \\
D3              &        0.33 &              0.67 &  0.5 &  0.5 &      1.0 &      1.0 &          1.0 &      1.0 &   0.14 & 0.56 & 1.0\\
QT-EWMA       &          0.83 &              0.87 &  1.0 &  1.0 &      1.0 &      1.0 &          1.0 &      1.0 &   1.0 & 1.0 & 1.0\\
TSDD     &    \textbf{0.0} &   \textbf{0.0} &  \textbf{0.0} &  \textbf{0.0} &      \textbf{0.0} &      \textbf{0.0} &          0.1 &      0.5 &   \textbf{0.0} & \textbf{0.0}  & \textbf{0.0} \\
ST &                \textbf{0.0} &              0.27 &  1.0 &  1.0 &      0.72 &      1.0 &          \textbf{0.0} &      \textbf{0.0} &   \textbf{0.0} & \textbf{0.0} & \textbf{0.0} \\
\hline

\end{tabular}
\end{table}

Table \ref{tabrealone} reports the detection results of models on real drift. 
A high true positive rate indicates a good detection by the model on real drift. 
QT-EWMA comes first with constant detection on all but one dataset. 
\pfirst~achieves perfect detection on five datasets, and above 50\% on four. \pfirst~performs well on the Musk, Insect and Waveform datasets which contain the most features.
\pfirst~comes in third just behind ADWIN which achieves a good detection on five datasets and some overall good scores across all but one dataset. 
The D3 model comes fourth yielding good results on over half of the datasets. The ST model makes four correct detections. TSDD performs poorly with no consistent detections. Both TSDD and ST have trouble detecting drifts on the datasets that contain the most features.

\begin{table}
\caption{Real Drift Detection}\label{tabrealone}
\centering
\begin{tabular}{lrrrrrrrrrrrrr}
\hline
 & Hyp. & Wav. &  Cov & Elec & Adult & Bank & D08 & Musk & Wine &    Ins.  & Chess & Lux. \\
\hline
\textbf{\pfirst}  &    \textbf{1.0} &   \textbf{1.0} &  \textbf{1.0} &  0.50 &  0.20  &   0.33 &  0.67 &     \textbf{1.0} &      \textbf{1.0} &   0.78 & 0.1 & 0.80\\
\hline
ADWIN     &   \textbf{1.0} &  0.10 &  0.60 &  0.50 &  \textbf{1.0}  &   \textbf{1.0} &         0.97 &         \textbf{1.0} &      \textbf{1.0} &   0.89 & 0.70 & \textbf{1.0} \\
D3              &                 \textbf{1.0} &               0.0 &  0.50 &  0.50 &  \textbf{1.0}   & 0.67 &         0.97 &        \textbf{1.0} &      \textbf{1.0} &   0.61 & 0.0 & \textbf{1.0}\\
QT-EWMA       &                 \textbf{1.0} &               0.90 &  \textbf{1.0} &  \textbf{1.0} &     \textbf{1.0}   &\textbf{1.0}  &         \textbf{1.0} &     \textbf{1.0} &      \textbf{1.0} &   \textbf{1.0} & \textbf{1.0} & \textbf{1.0}  \\
TSDD            &                 0.0 &               0.0 &  0.0 &  0.0 &     0.50  &  0.33 &         0.07 &     0.35 &      0.50 &   0.08 & 0.0 & 0.25\\
ST &                 0.0 &               \textbf{1.0} &  \textbf{1.0} &  0.30 &  0.35    &  \textbf{1.0} &         0.00 &        0.80 &      0.0 &   0.39 & 0.0 & 0.0 \\
\hline
\end{tabular}
\end{table}

In this benchmark, the drift detectors that ignore virtual drift, have trouble to detect real drift while the detectors that consistently detect real drift cannot ignore virtual drift. We demonstrated that \pfirst~raises little to no false alarms, while achieving a good level of performance in detecting real drift offering a good compromise in between a high false positive rate and a low true positive rate. \pfirst~is designed to work best when the input data follow Gaussian distributions but our experimental protocol shows that it is suitable when it is not the case.

\section{Conclusion}

The non discrimination of virtual and real drift is not suitable for industrial applications as it can generate a high number of false alarms, triggering costly label acquisition processes or long retraining times.
In this paper we introduced \pfirst, a novel drift detector that does not need labels to work. \pfirst~works best when the data distribution is normal by design. However, our experiment shows that it is able to detect drift even when the normality hypothesis is not respected. We demonstrated its ability to ignore virtual drift while keeping a good level of detection in the presence of real drift. The focus of future work will be adapting the algorithm to non Gaussian distributions as well as multi-class problems.

%
%
%
\bibliographystyle{splncs04}
\bibliography{biblio}

\begin{subappendices}
\renewcommand{\thesection}{\Alph{section}}%
\section{Demonstration of the quadratic values}

To compute the decision boundary $\alpha$, with $p_0$ and $p_1$ the proportion of samples of class 0 and 1. Let $F_0$ and $F_1$ be the density of each Gaussian distribution.

$$ F_{0}(x) = \frac{p_0}{\sqrt{2\pi}\sigma_0} e^{\frac{-(x-\mu_0)^2}{2\sigma_0^2}}$$

$$ F_{1}(x) = \frac{p_1}{\sqrt{2\pi}\sigma_1} e^{\frac{-(x-\mu_1)^2}{2\sigma_1^2}}$$

As $\alpha$ is the intersection of $F_0$ and $F_1$, when $x = \alpha$ we have:

$$F_{0}(x) = F_{1}(x)$$

$$\ln(F_{0}(x)) - \ln(F_{1}(x)) = 0$$

$$\ln(\frac{p_0}{\sqrt{2\pi}\sigma_0}) - \frac{(x-\mu_0)^2}{2\sigma_0^2} - \ln(\frac{p_1}{\sqrt{2\pi}\sigma_1}) + \frac{(x-\mu_1)^2}{2\sigma_1^2} = 0$$

$$\ln(\frac{p_0\sigma_1}{p_1\sigma_0}) + \frac{\mu_0^2}{2\sigma_0^2} - \frac{\mu_1^2}{2\sigma_1^2} + x (\frac{\mu_0}{\sigma_0^2} -\frac{\mu_1}{\sigma_1^2}) + x^2 (\frac{1}{2\sigma_0^2} + \frac{1}{2\sigma_1^2} )= 0$$

$\alpha$ can be computed analytically by solving the quadratic equation where the coefficients are:

$$a =  \frac{1}{2\sigma_0^2} - \frac{1}{2\sigma_1^2},\ b = \frac{\mu_1}{\sigma_1^2} - \frac{\mu_0}{\sigma_0^2},\ c = \frac{(\mu_0)^2}{2\sigma_0^2} - \frac{(\mu_1)^2}{2\sigma_1^2} - \ln \biggl(\frac{p_0\sigma_1}{p_1\sigma_0}\biggr)$$

Giving :

$$\alpha_0, \alpha_1 = \frac{\biggl( \frac{\mu_1}{\sigma_1^2} - \frac{\mu_0}{\sigma_0^2} \biggr) \pm \sqrt{\bigl(\frac{\mu_1}{\sigma_1^2} - \frac{\mu_0}{\sigma_0^2}\bigr)^2 - 4 \bigl(\frac{1}{2\sigma_0^2} - \frac{1}{2\sigma_1^2} \bigr) \bigl(\frac{\mu_0^2}{2\sigma_0^2} - \frac{\mu_1^2}{2\sigma_1^2} - \ln \bigl(\frac{p_0\sigma_1}{p_1\sigma_0}\bigr)\bigr)}}{2 \biggl( \frac{1}{2\sigma_0^2} - \frac{1}{2\sigma_1^2} \biggr)}$$

\end{subappendices}

\end{document}